**Title:** Trophic groups and modules: two levels of group detection in food webs

Benoit Gauzens[1,2*], Elisa Thébault[1], Gérard Lacroix[1,3] and Stéphane Legendre[4]

[1] UMR 7618-iEES Paris (CNRS, UPMC, UPEC, Paris Diderot, IRD, INRA), Université Pierre et Marie Curie, Bâtiment A, 7 quai St Bernard, 75252 Paris cedex 05, France

2 UMR 6553 Ecobio, université de Rennes 1, Avenue du Général Leclerc, Campus de Beaulieu, 35042 RENNES Cedex - France

[3] UMS 3194 (CNRS, ENS), CEREEP – Ecotron Ile De France, Ecole Normale Supérieure, 78 rue du Château, 77140 St-Pierre-lès-Nemours, France

[4] UMR 8197 IBENS (CNRS, ENS), École Normale Supérieure, 46, rue d'Ulm, 75230 Paris cedex 05, France

**Corresponding author *:**

Benoit Gauzens**:** benoit.gauzens@univ-rennes1.fr

**Short title:** Identifying trophic groups in food webs

**Abstract**

Within food webs, species can be partitioned into groups according to various criteria. Two notions have received particular attention: trophic groups, which have been used for decades in the ecological literature, and more recently, modules. The relationship between these two group concepts remains unknown in empirical food webs. While recent developments in network theory have led to efficient methods for detecting modules in food webs, the determination of trophic groups (groups of species that are functionally similar) is largely based on subjective expert knowledge. We develop a novel algorithm for trophic group detection. We apply this method to empirical food webs, and show that aggregation into trophic groups allows for the simplification of food webs while preserving their information content. Furthermore, we reveal a 2-level hierarchical structure where modules partition food webs into large bottom-top trophic pathways whereas trophic groups further partition these pathways into groups of species with similar trophic connections. This provides new perspectives for the study of dynamical and functional consequences of food-web structure, bridging topological and dynamical analysis. Trophic groups have a clear ecological meaning, and are found to provide a trade-off between network complexity and information loss.

## INTRODUCTION

In nature, species in communities are connected by their predation links, and these complex interactions can be represented by a network. The topology of these food webs is non-random and can have a considerable influence on their functionality [1,2], including their ability to persist. As for many complex networks [3], the notion of a group (a collection of nodes with specific characteristics) is a major topological feature of food webs [4–6], with important functional implications [7,8]. However this notion of group covers a large set of definitions (trophic groups, modules, regular equivalence groups, structural role groups, ... ) and methods (modularity maximisation, Markov chain clustering, statistical block modelling, spectral approaches, ...), giving different insights on network structure (see [9,10] for reviews on these notions). In food-web ecology, groups have been identified mainly according to two distinct definitions: modules and trophic groups (Fig. 1 A-B), but we still do not know how these two notions are related.

The notion of modularity (or community structure) refers to groups of nodes interacting more frequently between themselves than with other nodes. Modularity detection is challenging in view of its relation with network functionality [11]. For example, a modular structure can buffer the propagation of perturbations, determining the stability or resilience of ecological networks [8]. Mechanisms that give rise to modularity in food webs are not totally understood. Modules have been related to a variety of attributes, from niche organisation of species and their diet [12] to phylogeny [13] or spatial segregation between species [14]. For example, in the food web of Chesapeake Bay, the split found between two large modules corresponds closely to the division between pelagic and benthic species [15].

The study of food-web modularity is only recent, and historically, food webs have been mainly described in terms of trophic groups, in relation with the notion of trophic relationship introduced by Elton [16]. Trophic groups are constituted of species that share similar sets of preys and predators. Aggregation into trophic groups has been used to simplify the representation of food webs, circumventing methodological difficulties induced by the complexity of trophic relationships in empirical data [4,17], and allowing the comparison of datasets and models of similar resolution [18]. In fact, food webs were for a long time described at the trophic group level rather than at the species level [19,20]. The simplification of food webs into trophic groups is also central to the study of ecosystem dynamical and functional properties [21].

Several methods have been developed in order to detect trophic groups in food webs, based on two different notions. First, a set of methods inherited from the notion of *structural equivalence* [22]. Two nodes in a graph are said structurally equivalent if they relate to the same group of nodes. This assumption was then relaxed in order to allow nodes with similar but not identical relations to be said structurally equivalent. A classical method is to measure interaction similarity between nodes and then use a hierarchical clustering method (a stepwise classification process) to define structurally equivalent groups. In ecology, the Jaccard index has been used to define the amount of trophic overlap between taxa [23,24]. The main limit of the use of hierarchical clustering methods is that the number of groups does not appear as an emergent property, a threshold value for trophic similarity delimiting the groups or for the number of groups itself has to be preset.

A second way for detecting trophic groups in food webs is based on the notion of *regular equivalence,* inherited from the concept of role in social sciences [25]. A group of regularly equivalent nodes contains species that are connected to the same set of groups containing regularly equivalent nodes. Regular equivalence was introduced not to detect groups of nodes with similar interaction patterns but to aggregate entities with the same role. Regular equivalence is classically illustrated with the example of interactions in a hospital: two nurses do not necessarily interact with the same persons (they can have different patients, or interact with different doctors) but they interact with similar types of persons (patients, doctors...). Thus, nurses have the same role in the hospital The method of Luczkovich et al. [26] uses the notion of regular equivalence in ecology to group species, but the number of groups used for model selection has to be predefined and it potentially creates groups of species that do not share any trophic interactions. Block modelling approaches introduce an objective criterion for model selection. In their seminal paper, Allesina and Pascual [5] use AIC to select among models. In subsequent articles, Bayes Factors [14,27] or Normalized Maximum Likelihoods [28] were used. The main advantage of block modelling is the use of objective criteria for model selection, implying that the number of groups is not predefined. It however shares the same limit as all methods using the notion of regular equivalence by potentially aggregating nodes without any common connection (Fig 1C).

We propose here a new method of trophic group detection based on structural equivalence in order to avoid the limits of regular equivalence (lumping in the same group species without any common prey or predator), but with the ability to determine the number of groups as an emergent property of the system.

We then use the different notions of groups used in ecology to understand whether food webs are better described when grouped according to trophic groups or to modules, and whether modules and trophic groups give opposite, similar, or complementary descriptions of food-web topology. While modularity is gaining increasing interest in food-web studies [6,12,29], its relationship with trophic group arrangements is unknown as both network patterns have been studied independently. Detecting how different network decompositions are combined in food webs is important for understanding their structure and can reveal new network properties. It is also critical to assess the relevant and redundant features of network structure and to move beyond a disconnected view of food-web patterns. It has been shown that species aggregation into trophic groups did not affect the perception of food-web response to top-predator manipulation in an experiment [17]. Such result suggests that food webs might be mostly structured in trophic groups.

We therefore address here two different questions. First, we propose an efficient method to detect trophic groups in food webs. Second, using 9 aquatic food webs of different resolutions, we compare these trophic groups to groups obtained by modularity detection [15] and groups obtained by the model of Allesina and Pascual [5], thereafter referred as the AP model. The AP model is a block modelling approach that achieves the best compromise between the number of groups (network complexity) and information loss, using AIC for model selection. Depending on the structure of the considered network, the AP model will detect modules (i.e. groups of nodes interacting more frequently between themselves) or groups of regular equivalent species. The point is that classical methods for role detection create groups of regularly equivalent species (species in different groups are connected exactly to the same set of groups), whereas the AP method creates groups with group-specific connections to other groups. We show that trophic groups give a reliable picture of food webs in regard to information theory while preserving ecological significance, as we obtain close correspondences between the trophic group model and the AP model. This close matching does not hold when the methods are applied to two social networks, the Zachary's karate club [30] and the social prison inmate [31]. By comparing the trophic position of species in module and trophic group arrangements, we reveal a previously undetected link between trophic groups and modules: modules decompose the food web into disjoint vertical pathways of energy flow, and, within modules, trophic groups are composed of species of similar trophic levels.

**MATERIALS AND METHODS**

**A model for the detection of trophic groups**

A trophic group (TG) is usually defined as a group of species that interact with similar preys and predators. We mathematically translate this definition using the notion of trophic similarity [23] and the conceptual framework of modularity detection [32]. The notion of trophic similarity is related to the notion of structural equivalence. It allows to avoid the drawback of regular equivalence where species without any common interactions can be grouped in the same trophic group. Using comparison to a random model, modularity detection allows to obtain the number of groups as an emergent property, which is not possible when hierarchical classification methods are used to detect groups of structural equivalence.

The modularity of a given partition $E$ (a particular arrangement of the species in non-intersecting groups) in a network is given by the difference between the within-groups link density and its random expectation [33]:

$$M(E) = \sum_{s=1}^{|E|} \left( \frac{l_s}{L} - \left( \frac{d_s}{2L} \right)^2 \right), \tag{1}$$

where $|E|$ is the number of elements in the partition (the number of modules), $l_s$ is the number of links between nodes in the $s$ module, $L$ is the total number of links of the food web, and $d_s$ is the sum of degrees of species belonging to module $s$. The parameter $l_s/L$ is the fraction of links inside module $s$ (within-group link density), and $(d_s/2L)^2$ is an approximation of this expected quantity by chance alone.

For trophic group detection, we keep the comparison with a random null model, but instead of using the proportion of within-group links, our index is based on trophic similarity. The trophic similarity of two species is their number of common preys and predators divided by their total number of preys and predators. We transpose this definition using an analogy with the modularity index, by comparing the observed trophic similarity between all pairs of species in the same group to its expected value in a random graph. For a given partition $E$, our index is defined as

$$G(E) = \sum_{g=1}^{|E|} \frac{1}{|g|} \sum_{\substack{i,j \in G \\ i<j}} \left( T(i,j) - E(T(i,j)) \right), \quad (2)$$

where $|g|$ is the number of nodes in group $g$, $|E|$ is the number of groups in the partition $E$. $T(i,j)$ (and its expected value in a random graph $E(T(i,j))$) is the ratio between the number of preys and predators interacting with species $i$ and $j$, and the number of preys and predators interacting with species $i$ or species $j$:

$$T(i,j) = \frac{|P_i \cap P_j| + |p_i \cap p_j|}{|P_i \cup P_j| + |p_i \cup p_j|} \quad (3)$$

.

Here $P_i$ and $p_i$ represent respectively the set of predators and prey of species $i$, $|P_i \cap P_j|$ is the cardinality of the intersection of $P_i$ and $P_j$ (i.e., the number of prey and predators common to species $i$ and $j$). The value of $T(i,j)$ is directly obtained from the in- and out-degrees of species *i* and *j* in the food web. The computation of $E(T(i,j))$ is described in the Supplementary Information S1.

Group detection is performed by maximizing the trophic group index *G*(*E*) using a simulated annealing algorithm for each of the considered networks (Table 3). The N_W computer program was used to perform the computations [34].

**Networks studied**

Analyses were made on a dataset of 9 food webs and 2 social networks. Food webs were chosen for their low level of aggregation (i.e. most trophic interactions are described at species and genus level and not at the level of large trophic groups). The 9 food webs are: Benguela [35], Bridge Brooke Lake [36], Carribean Reef [37], Chesapeake Bay [38], Créteil Lake (Supplementary Information S3), Tuesday Lake [39], Carpinteria [40], DempsterSu [41], Ythan estuary [42]. The two social networks, the Prison inmate [31] and Zachary's karate club [30] graphs are classical examples in social science studies. They were used to assess whether the specific results we found for food webs were also relevant for other kinds of networks. A specific focus was put on the Lake Créteil food-web to investigate the characteristics of the trophic groups found by our method. The Lake Créteil food-web was created on the basis of a summer mesocosm study [4] conducted by G. Lacroix and colleagues; we thus have a good knowledge of the ecology of this food web.

**Comparison between group arrangements of the different detection methods**

In order to assess whether food webs are better described when grouped according to trophic groups or to modules, we compare the trophic groups obtained with our method and the modules to the groups obtained by the AP model. For both modularity and the AP model, we used a simulated annealing algorithm to detect groups in the considered food webs. To assess the correspondences between the different group detection methods, we used a mutual information criteria [33]. The normalised mutual information $I_{EF}$ between two partitions is defined as the ratio between the mutual information of the partitions and the mean of their respective entropy [43]:

$$I_{EF} = \frac{-\sum_{i=1}^{|E|}\sum_{j=1}^{|F|} n_{ij}^{EF} \log\left(\frac{n_{ij}^{EF} S}{N_i^E N_j^F}\right)}{\frac{1}{2} \times \left(\sum_{i=1}^{|E|} N_i^E \log\left(\frac{N_i^E}{S}\right) + \sum_{j=1}^{|F|} N_j^F \log\left(\frac{N_j^F}{S}\right)\right)} \quad (4)$$

Here, $S$ is the number of species, $|E|$ and $|F|$ are the number of groups in partitions $E$ and $F$ respectively, $N_i^E$ and $N_j^F$ are the number of nodes in group $i$ of partition $E$ and group $j$ of partition $F$. Finally, $n_{ij}^{EF}$ is the number of nodes that are both in group $i$ of partition $E$ and in group $j$ of partition $F$. The mutual information between partitions $E$ and $F$ is equal to 1 if both partitions are identical, and 0 if there is no matching.

**Relations between trophic groups and modules**

We investigated the links between trophic groups and modules in three ways: first by comparing the distribution of species trophic level between these two types of groups, second by measuring whether trophic groups were embedded in modules, and third by characterizing the contribution to modularity of species belonging to trophic groups that were split across different modules.

*1 - Distribution of species' trophic level in trophic groups and modules*

The trophic level of a species is defined as 1 plus the mean trophic level of its prey, with the trophic level of basal species set to 0. For all food webs, we calculated the variance in species trophic level either within modules or within trophic groups. To test whether variance of species trophic levels within modules differed from random expectation we used a null model approach. This null model distributes species randomly in different modules, whilst keeping the number of modules and their respective sizes as in the original network (100,000 replications, p-value is the probability to obtain a higher variance of trophic levels within the food web modules than expected from the null model). To test whether variance of species trophic levels within trophic groups differed from random expectation, we used the same null model as described above, but with a random attribution of species to trophic groups instead of modules (100,000 replications, in this case the p-value is the probability to obtain a lower variance of trophic levels within the trophic groups than expected from the null model).

*2- Module diversity of trophic groups*

To assess whether species affiliated to the same trophic group also belong to the same module, we measured an index of module diversity for trophic groups:

$$D_g = 1 - \sum_{s=1}^{m} \left(\frac{g_s}{|g|}\right)^2 \tag{5}$$

where $g_s$ is the number of species of group $g$ that belong to module $s$ (i.e., the cardinality of the intersection of $g$ and $s$), and $|g|$ is the number of species in $g$ (the underlying partition is implicit in this notation). $D_g$ is 0 if all species of a trophic group belong to the same module and is $1 - \frac{1}{|g|}$ when all species in the group belong to different modules. These values are compared to a null model where the partition into trophic groups is identical to that obtained with our model, but where species are randomly distributed among modules whilst keeping the same number of modules and their respective sizes as in the original food web. Comparisons are made with 100,000 values of diversity obtained with the null model.

*3 - Participation coefficient of species to modules*

We observed that each trophic group was in general embedded into a single module. We tested whether species of trophic groups that were split across different modules occupied a particular position within the modular structure. In order to determine the species contribution to network modularity, we computed the participation coefficient [44]. Based on the Simpson diversity index, the participation coefficient measures the diversity of connections of species $i$ to the different modules of the network:

$$PC_i = 1 - \sum_{S=1}^{m} \left(\frac{l_{is}}{d_i}\right)^2 \tag{6}$$

Here, $m$ is the number of modules, $l_{is}$ is the number of links between species $i$ and the species of module $s$, and $d_i$ is the degree (number of preys and predators) of species $i$. $PC_i$

$P_i$ equals 0 when all links of $i$ are in its own module, and is $1 - \frac{1}{m}$ when links are uniformly distributed among modules. Student's t- tests are then used to compare indices found for species in trophic groups belonging to different modules and species in trophic groups belonging to only one module.

### RESULTS

**The different aggregation methods are expected to return different groups**

This is shown using a simple network, a directed tree in which all species except the basal species have the same number of prey (Fig. 1). We can notice on Fig. 1 a major difference between AP groups and trophic groups: in the case of AP groups, all basal species are lumped together while it is not the case for trophic groups. With AP groups, species can belong to the same group even if they do not share any common predator (Table 1). In this particular topology, AP groups are equivalent to groups found using a regular equivalence method [26].

**Example of functional divisions in the food web of Lake Créteil**

In the food web of Lake Créteil, the trophic group (TG) method identifies 13 trophic groups. They tend to discriminate species according to trophic level (either phytoplankton, zooplankton, carnivorous or omnivorous) as well as body size, taxonomy and habitat (Table 2).

Note that it is difficult to assess the relevance of the group constituted of the trophospecies 'Bacteria', 'DOM and POM' (dissolved and particulate organic matter) and 'Biofilm', as the ecological role of these constituents can be different. This part of the network, which groups together detrital and littoral components of the food web, is not well known. Considering more precisely bacterial diversity and biofilm composition could lead to a different result.

Using module detection [33], we observe that most species within a trophic group belong to the same module (i.e., trophic groups are a sub-partition of modules, Fig. 2, Table 3). Thus, within a module, trophic groups interact mostly between themselves. Moreover, we can appreciate in Fig. 2B that modules assemble trophic groups along energetic pathways in the food web (vertical component). The first module (left part of Fig. 2B) brings together food chains involving small herbivorous zooplankton and Calanoids. The second module (middle part of Fig. 2B) brings together food chains involving large filter feeders (Cladocera). These two modules are mainly pelagic, and separate energetic pathways according to body size and behaviour of herbivores (small *vs* large graspers and filter feeders). The third module (right part of Fig. 2B) brings together trophic pathways dominated by organisms that are mainly omnivorous and are able to feed on littoral and benthic organisms. Hence, in the food web of Lake Créteil, modules appear as assemblages of trophic chains that link trophic groups with common major characteristics (size, behaviour, edibility, spatial niche).

**Comparison between group arrangements of the different detection methods**

In the 9 empirical food webs considered, TG always leads to partitions with a higher number of groups than modularity (Table 3). Indeed, modularity leads to partitions with a very low number of modules, suggesting that the number of independent subnets is limited (Table 3).

The number of groups obtained with the AP method is always higher than with modularity and lower than with TG (with one exception for the Benguela food web, Table 3).

Correspondence indices between groups obtained by TG and AP are significantly higher than correspondence indices between modularity and AP (paired Student's t-test, p<0.001). The high degree of overlap between TG and AP (Table 3, correspondence close to 1) suggests that an important part of the information carried by food-web structure can be attributed to trophic groups. Strikingly, and despite totally different goals, the AP method (looking for the most informative partitions) and the TG method lead to similar results (Table 3) even if the AP method still groups species without any common interaction whereas the TG method does not (Table 1). This close match between the two methods seems to be specific to food webs. Indeed, when comparisons are made on the two social networks, the Zachary's karate club and the prison inmate, correspondence indices are much lower with values of 0.531 and 0.478 respectively.

**Relations between trophic groups and modules**

*Distribution of species' trophic level in trophic groups and modules*

Food-web representations combining trophic levels of species and their affiliation to modules and trophic groups (Figs 2-3, Supplementary Information S2) suggest that, whereas species in the same trophic group tend to occupy the same trophic level, species in the same module often belong to different trophic levels. We computed the variance of species trophic levels within either modules or trophic groups. In all the food webs studied, the average variance of species trophic levels in modules was always higher than in trophic groups ($p<10^{-4}$ for all networks). Furthermore, the variance of trophic levels of species belonging to the same module was higher to what is expected by chance alone ($p<10^{-4}$ for all food webs). The opposite pattern was found when considering the variance of trophic levels of species sharing the same trophic groups ($p<10^{-4}$ for all networks). By definition, species in a module are highly connected. As most trophic relations occur between species of different trophic levels, this could explain why species in the same module tend to belong to different trophic levels. Therefore, modules reflect particular energetic pathways, representing parallel trophic chains.

*Modules diversiy of trophic groups and participation coefficient of species to modules*

We observe that species in a trophic group tend to belong to a same module (Figs 2-3, Supplementary Information S2). Thus, trophic groups tend to be embedded in modules. For all food webs, the average module diversity $D_g$ of trophic groups was close to 0 and belonged to the 5% lowest values generated from the null model. This highlights a hierarchical two-level structure of food webs, where a partition into modules is further partitioned into trophic groups.

Although striking, this arrangement of trophic groups into modules is not perfect. Species of a given trophic group are in some instances dispatched in different modules. The mean participation coefficients to modules of species in trophic groups dispatched in different modules are in most cases significantly lower than species in groups that belong to a single module. Indeed, in most food webs, the species of trophic groups that are split in several modules are those that contribute the least to the modular structure of the food webs (Table 3).

**DISCUSSION**

Thanks to the development of a new algorithm to identify trophic groups in food webs, our study reveals two important features of the structure of empirical food webs. First, we show that lumping species according to trophic groups allows the simplification of food webs while preserving the information carried by the initial network structure. Second, by considering together trophic groups and modules, we put forward a previously unnoticed pattern of organisation of food webs: modules are composed of species from different trophic levels, and are further partitioned into trophic groups; they represent energetic pathways linking trophic groups from the bottom to the top of the food web.

**An algorithm to identify trophic groups**

Whereas the concept of trophic group is widely used in the ecological literature since Elton [16] and Lindeman [45], the characterization of trophic groups is usually based on (subjective) expert knowledge. In the existing methods of food-web aggregation into trophic groups [23,24], the number of trophic groups is defined by the user and is not an emergent property of the network. Using the methodology developed for modularity indices, our method of trophic group detection circumvents previous limitations [5,46] where the ecological meaning of the partitions returned does not come from the method itself. By contrast, our method is based on the ecological notion of trophic similarity, and by extension on the notion of nodes with similar patterns of connections.

**Trophic groups: main underlying structure of food webs?**

The trophic group method and the AP method detect groups according to totally different criteria. The AP method aims at finding partitions corresponding to the best trade off between information loss and reduction of complexity using the AIC, without any notion of ecology. The trophic group method finds clusters of species with similar sets of prey and predators. The match found between the partitions returned by the two methods shows that trophic groups support a large part of the information carried by the underlying structure of the food web, as given by the AP method. The relevance of species aggregation into trophic groups has already been suggested to reflect functional properties [4,17,21,47] or to identify structural patterns [14]. We highlight here that food-web decomposition into trophic groups aggregates species with minimal loss of information while keeping a clear ecological meaning, and with the potential to reflect the functioning of the network. The relevance of such aggregation criteria (groups of nodes interacting with similar groups of nodes) seems very general for food webs. On the other hand, the aggregation process did not prevent information loss when it was applied to the two social networks. An intuitive explanation might be that species with similar prey and predators do not predate on each other while in social networks, actors with similar relationships tend to know each other and are often not precluded from interacting.

**Trophic groups and modules: complementary views of food-web structure**

Though we show that the notion of trophic group prevails in food webs, our study also confirms that modules are an important feature. Previous studies have already shown that food webs are more modular than random networks [12]. This suggests patterns of organization similar to those observed in other biological networks (gene-protein, plant-pollinator, neuronal), and in some small-world networks [48]. While modular patterns still need to be explained in food webs, we observe that modules represent parallel pathways of energy from producers to consumers, delimiting distinct food chains (Figs. 2-3, Supplementary Information S2). This is in accordance with previous results [12] showing that species in the same module (according to the notion of directed modularity) are globally located on trophic chains coming from similar basal species. We reveal that the variance of species trophic levels within modules is higher than expected by chance. The opposite result is found when groups are determined only accordingly to prey or predator similarity [12].

Despite having intuitively nearly opposite definitions (modules represent groups of species interacting mostly with one another whereas trophic groups correspond to groups of species interacting with other well-defined groups of species), modules and trophic groups are linked and provide complementary pictures of food-web structure. It appears that food webs present a two-level hierarchical structure, with each trophic group belonging globally to a single module. The existence of network hierarchical structure has already been described for social networks [49]. Some trophic groups are however sometimes split across several modules. Species of such trophic groups share the same neighbourhood, as they are in the same trophic group, but belong to different communities (modules). These species are connected more diversely to modules than other species, therefore, they potentially bridge different modules. As the modular structure limits the propagation of perturbations [8], species bridging different modules could play a key role by interconnecting distinct subnets of energetic pathways, and allowing different ecological processes (perturbations, trophic cascades, ...) to shift from a module to another.

**Implications for future research**

The functional implications of modularity are currently widely explored [8,50], but little is known about the functional implications of the trophic group structure. Indeed, while modules are characterized by a high density of within links, the implications of the architecture defined by trophic groups (few links within trophic groups, and a large number of links between some trophic groups) have not been addressed. Trophic groups are often used as a simplification, making the system more readable, sometimes as a consequence of external constraints (spatial segregation [14]), but the functional implications of trophic group patterns are worth exploring. For example, we still do not know how the dynamics of trophic groups is related to the individual dynamics of their component species.

Species richness within trophic groups could be considered as functional redundancy. The deletion of a whole group might lead to the loss of an entire set of specific connections, which could potentially have dramatic effects on system properties. As many topological studies [51–54] focus on the detection of key species in networks, the determination of the aggregated network of trophic groups addresses the question in a new way by considering potential key species as elements of trophic groups characterized by a low diversity.

As food-web descriptions are becoming more and more precise — recent published food webs contain several thousands of links — the reduction of complexity will become a critical issue. Our approach has the advantage of delineating trophic groups in such a way that complexity is reduced while keeping a clear ecological meaning. However, we need to know the entire network to simplify it. The next step will be to consider the correspondences between the biological traits of species within and between trophic groups, in order to develop methods able to reconstruct trophic groups and their links using species attributes. Addressing this question may improve our comprehension of the parameters involved in the trophic niche space (set of ecological parameters determining the trophic relationships of species). Several parameters, such as size [55], phylogenetic relationships [13,27], or behaviour [56] have been already considered. Even if they are limited to trophic relationships, these studies might provide a useful tool for the generic classification of species.

Improving our comprehension of network simplification is essential to address the structure-function relationship in food webs. As modelling approaches cannot encompass the entire complexity of food webs, food-web simplification via trophic group detection provides a trade-off between consistency and mathematical tractability, relating structural properties and functional issues.

**Acknowledgements**

We are grateful to Stefano Allesina for his review of the paper and his comments. BG was supported by a grant from the R2DS Ile-de-France program (project no. 2007-14) and acknowledges the EC FP7 FET support. SL, GL, and ET acknowledge support from the Agence Nationale de la Recherche ("ANR BLANC" PHYTBACK project, "ANR CEP&S", PULSE project)." The funders had no role in study design, data collection and analysis, decision to publish, or preparation of the manuscript.

Table 1: The number of pairs of species belonging to the same groups but without any common interactions is nonzero for the AP method and almost zero for the trophic group method.

| Networks | AP | Trophic groups |
|---|---|---|
| Creteil | 0 | 0 |
| DempsterSU | 73 | 0 |
| Tuesday Lake | 11 | 0 |
| Cheasapeake Bay | 62 | 0 |
| Ythan Estuary | 62 | 0 |
| Bridge brook lake | 7 | 0 |
| Caribean reef | 27 | 1 |
| Carpinteria | 39 | 1 |
| Tuesday lake | 11 | 0 |

Table 2. Groups obtained by our trophic group detection method (left) in relation to group characteristics (right) for the Lake Créteil food web These groups are represented by the corresponding colours in Fig. 2.

| Trophic groups | Group characteristics |
|---|---|
| *Abramis brama, Rutilus rutilus, Acanthocyclops robustus* | Omnivorous fish and large Cyclopoids (blue-green) |
| *Asplanchna girodi, Asplanchna priodonta, Thermocyclops crassus, Thermocyclops oithonoides* | Carnivorous Rotifers and small Cyclopoids (white) |
| *Eudiaptomus gracilis, Eutytemora velox* | Omnivorous Calanoids (green) |
| *Cephallodella* sp., *Chydorus sphaericus, Lecane bulla, Lecane luna, Lecane stichaea, Lepadella* sp., *Testidunella patina,* Chironomidae | Benthic or littoral species and detritivorous or bactivorous organisms (brown) |
| *Hexarthra mira, Filinia longiseta* | Rotifers consuming small algal cells and bacteria (pink) |
| Bdelloid species, *Bosmina coregoni, Bosmina longirostris, Brachionus angularis, Brachionus calyciflorus, Brachionus quadridentatus, Keratella cochlearis, Keratella quadrata,* nauplii of calanoïda, nauplii of cyclopidae, *Polyarthra dolichoptera-vulgaris, Polyarthra major, Pompholyx sulcata, Trichocerca* sp. | Small herbivorous zooplankton (dark green) |
| *Ceriodaphnia dubia, Ceriodaphnia pulchella, Daphnia cucullata, Daphnia galeata, Daphnia galeata x D. cucullata, Diaphanosoma brachyurum, Synchaeta pectinata* | Large herbivorous Cladocera (purple) |
| DOM and POM, Bacteria, Biofilm | Components of the detrital and littoral pathway (orange) |
| *Ceratium hirundinella, Nitzschia* sp., *Pediastrum boryanum, Synedra ulna* | Large or protected, poorly edible, algae (light purple) |
| *Dictyosphaerium pulchellum, Navicula* sp., *Pediastrum duplex, Schroederia indica, Staurastrum* sp., *Trachelomonas* sp. | Algae mainly consumed by graspers within zooplankton (light blue) |
| *Coelastrum* spp, *Colacium* sp., *Cosmarium* sp., *Cryptomonas* sp., *Desmodesmus quadricauda, Oocystis lacustris, Scenedesmus acuminatus* | Edible algae consumed by herbivorous and omnivorous zooplankton (dark blue) |
| *Chroomonas* sp., *Crucigenia* spp, *Cyclotella ocellata, Monoraphidium contortum, Tetraedron minimum* | Edible algae consumed by herbivorous zooplankton (yellow) |
| *Quadricoccus ellipticus,* small undetermined unicells | Small phytoplanktonic species, highly Edible algae for filter feeders (red) |

Table 3. Number of groups obtained using trophic groups (TG), modularity (M), and the Allesina & Pascual (AP) detection methods, with the degree of overlap between the different partitions. P sets the p-value of the difference of participation coefficients between species in trophic groups belonging to different modules and species in trophic groups belonging to only one module. D is the p-value of the difference in diversity of modules for trophic groups compared with a null model. The star symbol corresponds to food webs for which all trophic groups are in a single module. Hence, statistical analyses on P were not relevant in this case.

| | species (links) | TG | AP | M | TG-AP overlap | Module-AP overlap | P | D |
|---|---|---|---|---|---|---|---|---|
| **Benguala** [35] | 29 (203) | 7 | 7 | 3 | 0.841 | 0.397 | 0.0459 | $<10^{-4}$ |
| **Bridge Brooke Lake** [36] | 75 (553) | 12 | 9 | 3 | 0.92 | 0.631 | * | $<10^{-4}$ |
| **Carribean Reef** [37] | 249 (3313) | 46 | 28 | 3 | 0.775 | 0.365 | $<10^{-4}$ | $<10^{-4}$ |
| **Chesapeake Bay** [38] | 33 (72) | 13 | 7 | 3 | 0.745 | 0.428 | 0.4793 | $<10^{-4}$ |
| **Créteil Lake SI3** | 67 (718) | 13 | 12 | 3 | 0.922 | 0.4738 | 0.0194 | $<10^{-4}$ |
| **Tuesday Lake** [57] | 73 (410) | 17 | 11 | 2 | 0.834 | 0.449 | * | $<10^{-4}$ |
| **Carpinteria** [40] | 128 (2290) | 37 | 28 | 3 | 0.872 | 0.379 | 0.289 | $<10^{-4}$ |
| **DempsterSu** [41] | 107 (966) | 25 | 12 | 3 | 0.7129 | 0.410 | $<10^{-4}$ | $<10^{-4}$ |
| **Ythan estuary** [42] | 92 (409) | 26 | 13 | 3 | 0.755 | 0.317 | $<10^{-4}$ | $<10^{-4}$ |

Figure 1. Representation of different group detection methods for a hypothetical food web: A) modularity (3 modules), B) trophic groups method (5 trophic groups), and C) AP method (3 AP groups). Nodes of the same colour and with the same numbers belong to the same group. This hypothetical food web has the topology of a $N$-levels tree where each non basal species has exactly $d$ prey. Different partitions of this example of food web ($N=3$, $d=3$, $S=13$ species) are shown: 3 modules, 5 trophic groups, and 3 AP groups. In the general case of a regular $N$-levels directed tree with in-degree $d$, the number of species is $S = 1 + d + \cdots + d^{N-1}$. The number of modules, trophic groups, and AP groups are respectively $d$, $1 + 1 + d + \cdots + d^{N-2}$, and $N$. These numbers differ in general, with more trophic groups than modules or AP groups. We can observe here that AP groups correspond in this case to regular groups, based on the regular equivalence definition.

Figure 2. Representation of the Lake Créteil food web partitioned with the trophic group method (A,B), and module detection (B). In (A), trophic groups are delimited by coloured discs whose sizes are proportional to the number of species in each trophic group, and species are represented by small grey circles. In (B) modules are delimited by grey rectangles, and species are represented by small circles whose colour corresponds to their trophic group in (A). The vertical dimension corresponds to the species' trophic levels (B) and the average trophic level of trophic groups (A). The compositions and characteristics of the trophic groups for the Lake Créteil food web are described in Table 2.

Figure 3. Representation of the Tuesday Lake (A,B), DempsterSu (C,D) and Ythan Estuary (E,F) food webs, with species sorted according to their trophic groups (A,C,E) and their modules (B,D,F). Same conventions as in Fig. 2.

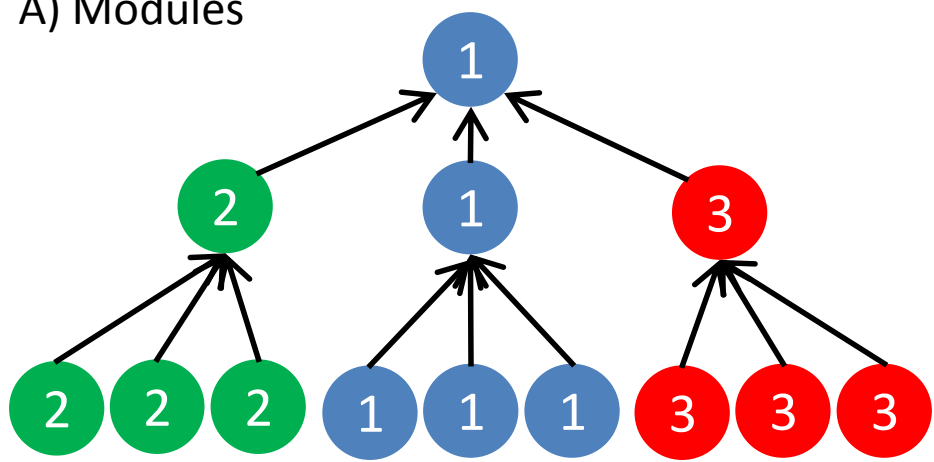
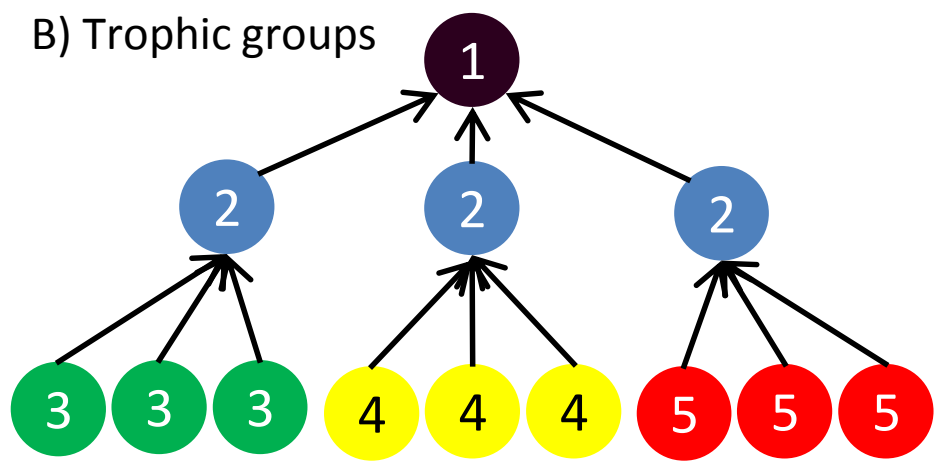
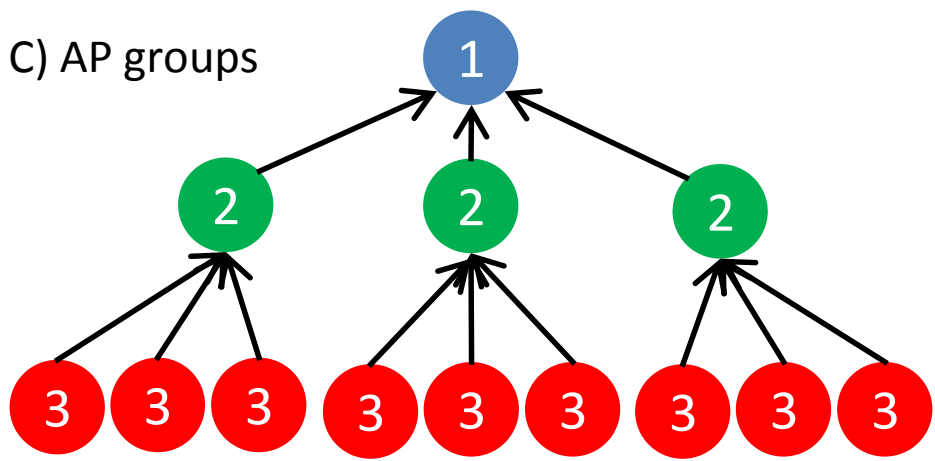

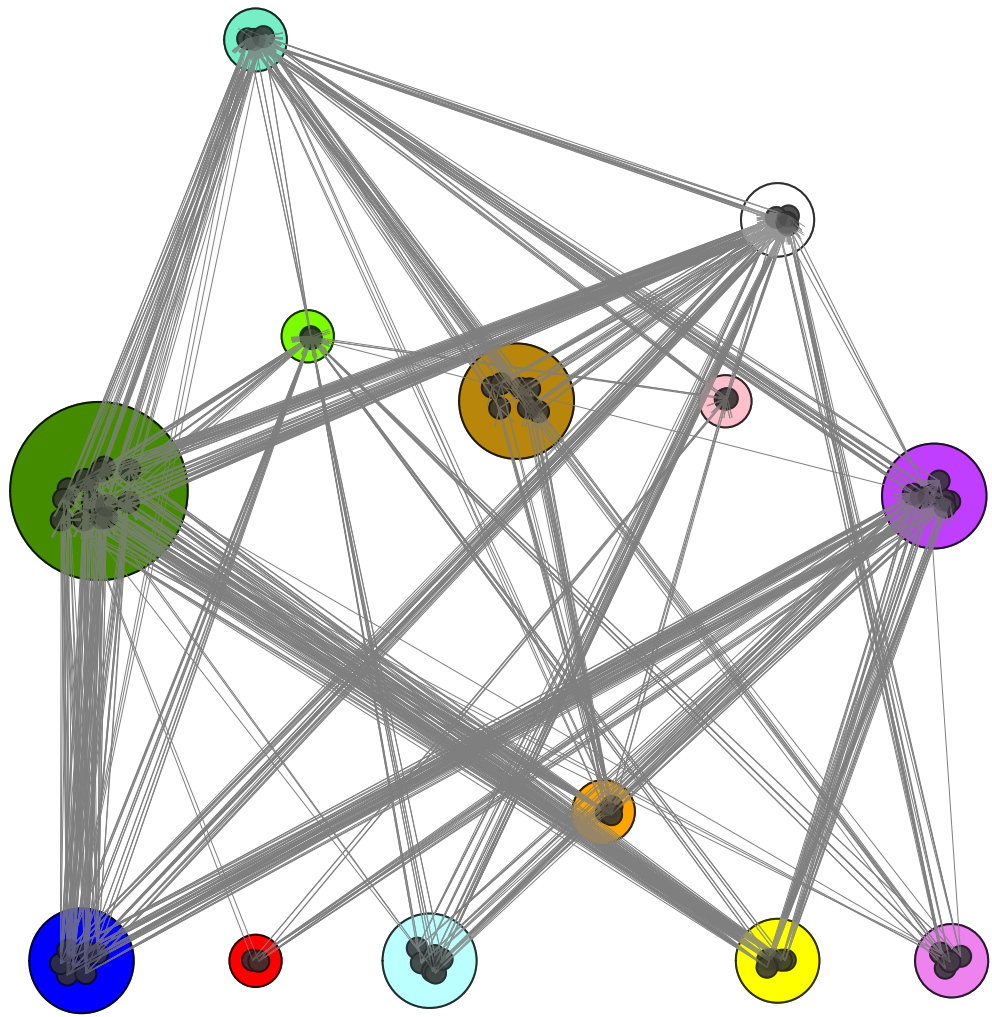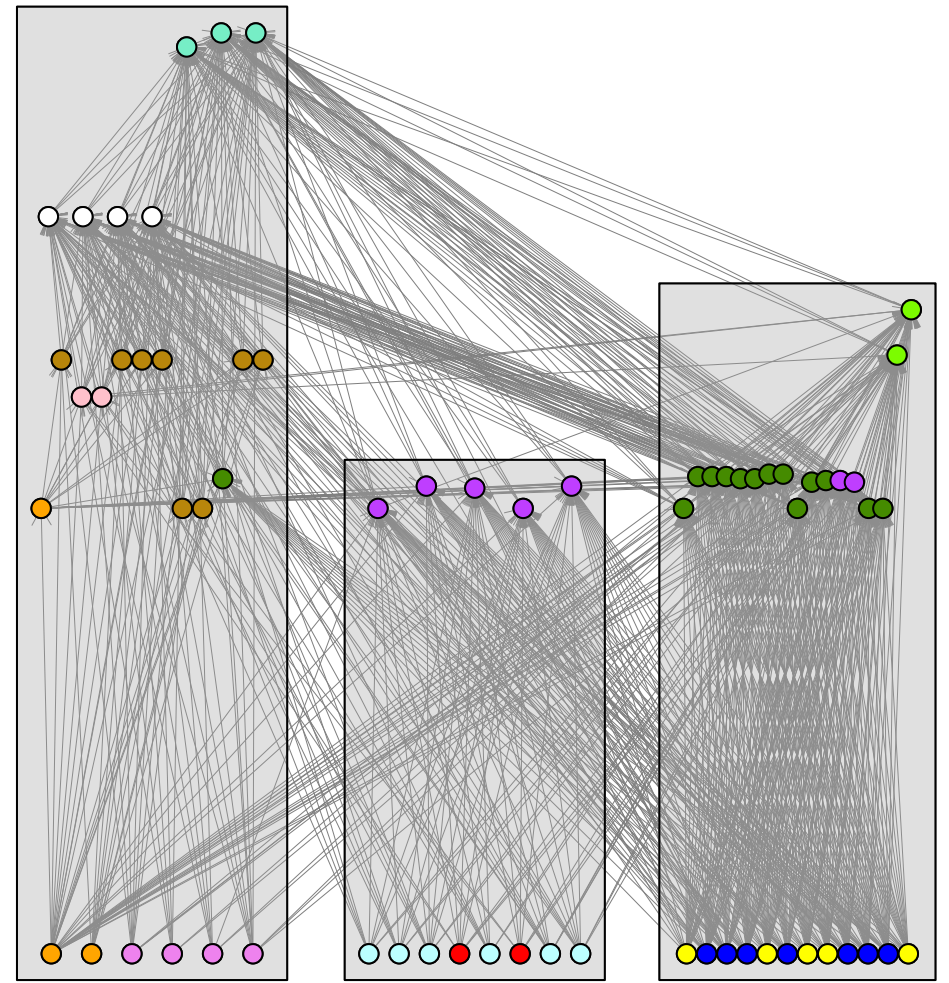

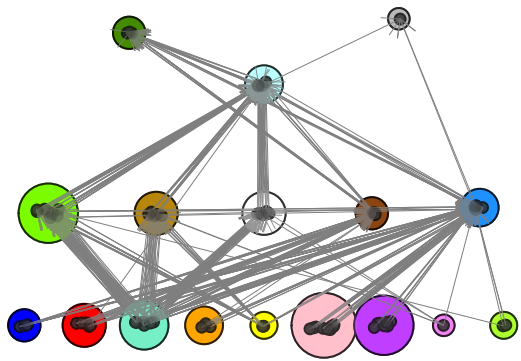 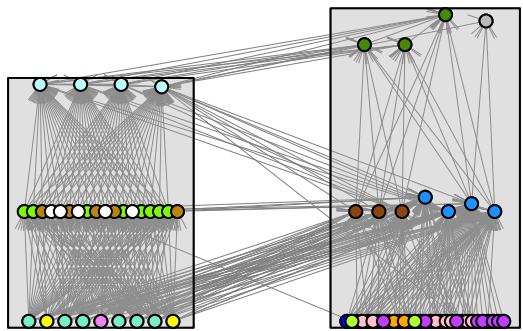
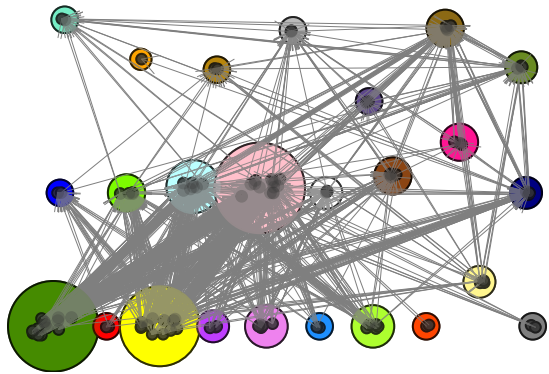 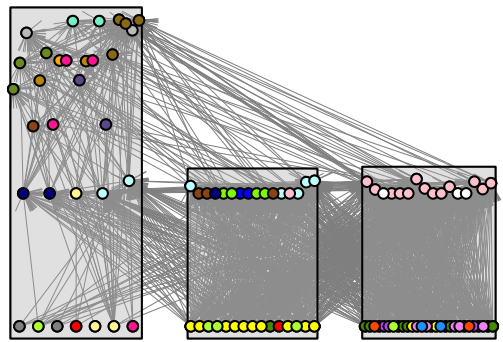
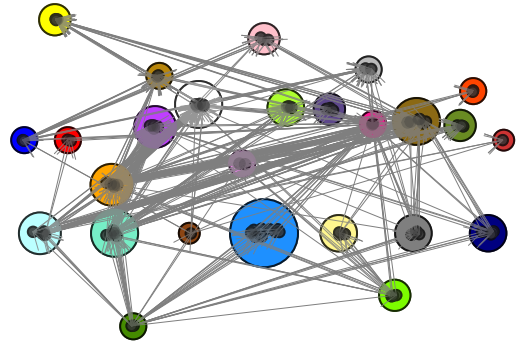 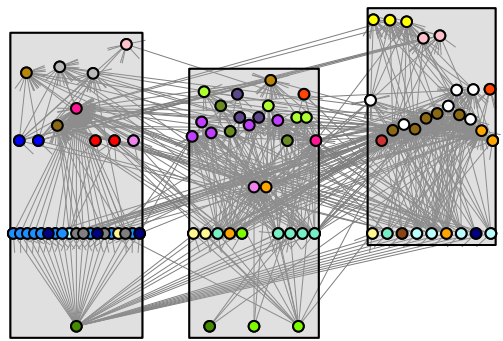